# A2AS FRAME WORK

## Agentic AI Runtime Security and Self-Defense

AUTHORS AND CONTRIBUTORS TO THE A2AS FRAMEWORK

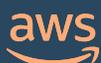 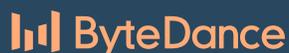 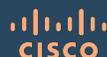 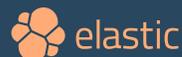 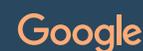

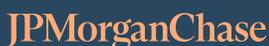 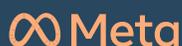 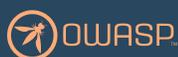 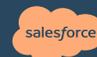 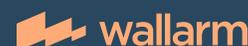

# Table of Contents

# A2AS: Agentic AI Runtime Security and Self-Defense

## Authors

**Eugene Neelou** [1, 9, 11], **Ivan Novikov** [1, 11], **Max Moroz** [1, 3], **Om Narayan** [2, 9], **Tiffany Saade** [4], **Mika Ayenson** [5], **Ilya Kabanov** [6], **Jen Ozmen** [6], **Edward Lee** [7], **Vineeth Sai Narajala** [8, 9], **Emmanuel Guilherme Junior** [9], **Ken Huang** [9], **Huseyin Gulsin** [9], **Jason Ross** [10], **Marat Vyshegorodtsev** [10], **Adelin Travers** [12], **Idan Habler** [12], **Rahul Jadav** [12]

| | | | |
|---|---|---|---|
| [1] A2AS | [2] AWS | [3] ByteDance | [4] Cisco |
| [5] Elastic | [6] Google | [7] JPMorganChase | [8] Meta |
| [9] OWASP | [10] Salesforce | [11] Wallarm | [12] Other |

## Abstract

The A2AS framework is introduced as a security layer for AI agents and LLM-powered applications, similar to how HTTPS secures HTTP.

A2AS enforces certified behavior, activates model self-defense, and ensures context window integrity. It defines security boundaries, authenticates prompts, applies security rules and custom policies, and controls agentic behavior, enabling a defense-in-depth strategy.

The A2AS framework avoids latency overhead, external dependencies, architectural changes, model retraining, and operational complexity.

The BASIC security model is introduced as the A2AS foundation:

(B) Behavior certificates enable behavior enforcement,
(A) Authenticated prompts enable context window integrity,
(S) Security boundaries enable untrusted input isolation,
(I) In-context defenses enable secure model reasoning,
(C) Codified policies enable application-specific rules.

This first paper in the series introduces the BASIC security model and the A2AS framework, exploring their potential toward establishing the A2AS industry standard.

# 1 Introduction

## 1.1 Traditional AI Security

The advancements in Artificial Intelligence (AI) and its integration across sensitive fields, such as healthcare, finance, and critical infrastructure, have increased the attack surface of such AI systems. They expose applications and data to risks of exfiltration, infection, and manipulation, potentially compromising confidentiality, integrity, and availability. Moving beyond theoretical risks, a growing number of real-world AI security incidents are being reported [1].

## 1.2 Generative AI Security

The developments in Large Language Models (LLMs) have introduced a paradigm shift in AI engineering, where building AI systems is largely centered around integrating LLM models. These models have their inherent vulnerabilities that expand the attack surface and introduce additional security risks. Real-world incidents include data breaches, compromised behaviors, and bypassed safety restrictions [2].

## 1.3 Agentic AI Security

The emerging agentic AI paradigm relies increasingly on LLM models for reasoning and task planning. Beyond inheriting all vulnerabilities of the underlying LLM models, AI agents introduce their own attack surface through task execution, tool usage, and protocol interactions. These factors make agentic AI systems vulnerable by design, requiring deliberate security hardening. As a growing number of organizations deploy and integrate AI agents with internal systems, security risks scale from isolated failures to systemic enterprise-wide incidents [3].

## 1.4 Prompt Injection

LLM models process external data and system instructions within a unified context window. While this feature enables model reasoning, it creates a critical vulnerability because trusted internal instructions and untrusted external inputs coexist in the same context window without clear security boundaries [4].

Prompt injection represents an emerging class of attack techniques exploiting this vulnerability. Attackers can inject malicious instructions to subvert intended model behavior, enabling a variety of attacks that lead to agent manipulation or data exfiltration [5].

## 1.5 Scope of Work

In this paper, we make several contributions.

First, we introduce the **BASIC SECURITY MODEL**, which defines a set of essential security primitives for agentic AI runtime security.

The BASIC security model covers behavior certification, context window integrity, and secure model reasoning with controls such as behavior certificates, authenticated prompts, security boundaries, in-context defenses, and codified policies.

Second, we introduce the **A2AS FRAMEWORK**, an implementation of the BASIC model, that serves as a runtime security layer for AI agents and LLM-powered applications, similar to how HTTPS secures HTTP.

A2AS enforces certified behavior, activates model self-defense, and ensures context window integrity. It defines security boundaries, authenticates prompts, applies security rules and custom policies, and controls agentic behavior, enabling a defense-in-depth strategy.

The framework provides an efficient solution, operating at runtime and within the native context window, delivering effective protection and enabling agentic AI security at scale.

The following sections review existing limitations, introduce the BASIC security model and the A2AS framework, explore A2AS use cases, and outline a roadmap toward establishing the A2AS industry standard.

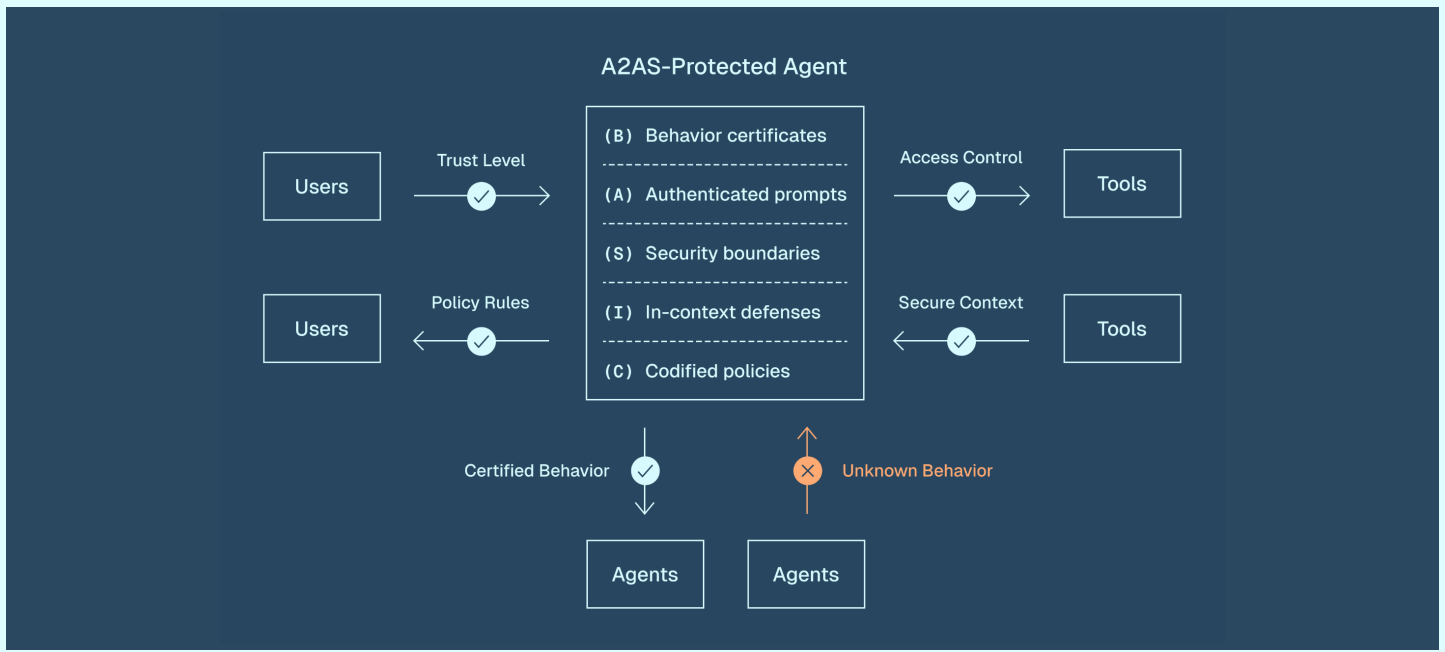

**Figure 1:** A2AS-protected AI agent with BASIC security controls

# 2 Problem Definition

## 2.1 Intrinsic Vulnerability

Current neural networks have intrinsic vulnerabilities that enable security attacks such as adversarial examples and prompt injections.

The first attacks against machine learning algorithms were published over twenty years ago [6]. To date, solutions with security guarantees for adversarial examples that exploit the decision boundaries of neural networks still have not been found [7].

Similarly, prompt injections exploit the intrinsic LLM vulnerability. They exist due to the way LLM models process external inputs, which places trusted internal instructions and untrusted external data in the same context window without clear security boundaries.

Attackers take advantage of this LLM design vulnerability to introduce malicious commands into the context, instructing models and agents to deviate from intended behaviors and perform unauthorized actions. Common examples include direct prompt injections, where attackers append explicit malicious instructions to inputs, and indirect prompt injections, where malicious content is embedded within inputs from external sources such as documents or web URLs [8].

AI agents and LLM-powered applications inherit vulnerabilities from the design of LLM models, expanding the attack surface and making them vulnerable to attacks that cannot be fully eliminated by design.

## 2.2 Ineffective Defenses

The range and complexity of LLM attacks are evolving rapidly, driven by the adoption of AI systems. Deployment in mission-critical domains motivates the development of prompt injection defenses. However, existing AI security solutions lack universal effectiveness [9].

**HEURISTIC** methods rely on pattern matching and keyword filtering to block known malicious payloads. However, these basic guardrails fail to detect advanced or novel prompt injection techniques due to their reliance on static patterns [10].

**DETECTION** methods, although easy to implement, are also easy to bypass with novel attack techniques. Additionally, the design of these guardrails involves external content classification models that often introduce significant or even prohibitive latency [11].

**SEMANTIC** methods involve embedding explicit security instructions into prompts to guide model behaviors. While demonstrating effective prevention, these methods depend on the model's own reasoning for compliance with instructions, which makes consistent protection across diverse LLM model versions complicated [12].

**DESIGN** methods use multi-model architectural isolation and sandbox environments to process and validate external inputs separately from the primary model execution. Although theoretically robust, these methods introduce significant technical complexity, making them impractical in many AI deployment scenarios [13].

Collectively, these methods have fundamental limitations and lack security guarantees. This unreliable effectiveness shifts the focus from finding the best defenses to managing unavoidable tradeoffs.

## 2.3 Impractical Complexity

A variety of prompt injection defenses have been proposed. However, their implementation introduces critical tradeoffs such as increased latency, external dependencies, architectural changes, token overhead, model retraining costs, and degraded AI performance.

**SECURITY-ALIGNED TRAINING** methods such as SecAlign rely on fine-tuning LLM models on mixed adversarial and benign prompts, so the model learns to refuse malicious instructions during inference [14]. This built-in defense strategy leverages the model reasoning without introducing extra latency and token costs. However, training secure models requires expensive compute resources and must be regularly repeated to keep up with emerging attacks. Additionally, AI engineers integrating third-party LLM models cannot benefit from it.

**CAPABILITY-CONTROLLED SANDBOXING** methods such as CaMeL use privileged and quarantined LLM models along with action planning and security policies [15]. Prompt inputs go through extraction control flow and capability checks, introducing extra latency and token costs. This approach requires restructuring the application for enforcement flow, which may be impractical in many AI deployment scenarios.

**INFORMATION FLOW CONTROL** methods such as F-secure force LLM models to make a structured plan, which is inspected before execution so that models process only validated requests [16]. Initial latency is low, but it grows with the plan depth. This also requires adapting the application to the control flow paradigm.

**CLASSIFICATION-BASED GUARDRAILS** such as GenTel-Safe place a dedicated protection model in front of the target LLM model to allow or block prompts [17]. Attack detection inference adds significant latency, and protection effectiveness can decrease over time if these guardrails are not frequently updated to detect emerging attacks.

**MULTI-LAYERED GUARDRAIL SYSTEMS** such as LlamaFirewall offer a modular framework with multiple detectors for security and alignment [18]. Complex orchestration requires a non-trivial deployment strategy. The multi-call design introduces significant latency due to sequential model execution, routing overhead, and control flows.

Ultimately, these methods introduce critical tradeoffs and increased operational complexity. Such impractical solutions leave a clear need for an effective and efficient enterprise-grade AI security solution.

# 3 Relevant Work

## 3.1 Protocol Vulnerabilities

The progress in agentic AI system development has introduced various interaction protocols between users, agents, tools, and environments. Besides providing opportunities for connecting AI systems, protocols such as A2A and MCP have further expanded their attack surface [19].

Threat modeling of the **MCP PROTOCOL** identified a wide range of security risks for agentic AI systems, including tool poisoning and MCP server impersonation [20]. Analysis of thousands of MCP servers revealed architectural vulnerabilities and highlighted security and maintainability problems [21]. Proposed MCP security practices focus on securing traffic, sanitizing inputs, and verifying identities [22].

Threat modeling of the **A2A PROTOCOL** proposed ways to address risks in multi-agent communications, including message integrity and agent authentication [23]. Proposed A2A security improvements focus on protecting data exchange, validating agent identities, and reducing the risk of agentic behavior manipulation [24].

## 3.2 Prompt Injections

A systematic study of LLM security in agentic AI systems shows that **PROMPT INJECTION** enables a wide range of attack scenarios. These attacks can go beyond altering outputs and bypass system policies, misuse trusted tools, or leak sensitive data [25].

In multi-agent systems, malicious instructions can propagate between AI agents as **PROMPT INFECTION**, creating persistence across tasks [26]. This enables attacks where a malicious prompt can cascade into denial-of-service, unauthorized actions, and coordinated attacks.

Prompt injections stand out as a core challenge for agentic AI security, affecting users, agents, tools, and environments while enabling a wide range of attacks. Addressing LLM security risks requires an approach that protects both AI agent communications and behaviors.

## 3.3 Security Instructions

Defenses increasingly use context-level methods to leverage context window augmentations for activating secure model reasoning.

**INSTRUCTION** defenses suggest to prepend or append prompts with inline reminders for LLM models about their primary task [27].

**FORMATTING** defenses introduce delimiters and tags to wrap inputs and help LLM models separate internal from external instructions [27].

**SPOTLIGHTING** offers several strategies that combine instruction and formatting methods such as delimiting, datamarking, and encoding to improve protection against indirect prompt injection attacks [28].

**BOUNDARY AWARENESS** approach introduces border strings for inputs and applies in-context learning to teach LLM models the boundaries between data and instructions. However, the core idea is based on teaching models to behave, not on security hardening [29].

**FORMATTING AUTHENTICATION** uses a hierarchy of tags for layers of external inputs and system instructions, enforcing tag-based rules. While promising, it requires custom rules for every AI application [30].

**STRUCTURED QUERIES** approach proposes a fine-tuning process that teaches LLM models to only follow structured instructions. They require custom encoders to convert external inputs into a custom format that a newly trained LLM model learned to recognize [31].

**DEFENSIVE TOKENS** approach suggests training LLM models with security instructions, extracting security embeddings, and applying them at inference time. While this method allows activating protection on demand for selected prompts, it provides no policy control [32].

**INSTRUCTION HIERARCHY** approach proposes an alignment strategy that teaches LLM models to recognize different priority levels for internal system instructions and external inputs. This method requires extensive security data generation and model retraining [33].

The reviewed context-level defenses show promising performance, but some lack effectiveness while others remain research concepts. A2AS aims to use the best ideas with enterprise-grade security engineering.

## 3.4 Behavior Enforcement

Recent research explored behavior-focused methods for agentic AI security such as runtime policies and capability specifications to manage agentic behaviors, tool usage, and access control.

**GUARDAGENT** uses a guardrail agent that generates safe execution plans based on specified system rules and access control policies. This approach relies on manual policy specification and requires the use of an external reasoning model, limiting its practicality [34].

**AGENTSPEC** uses a domain-specific language and a tool for behavior enforcement, intercepting execution flow within agentic frameworks. The approach focuses on business logic rather than security [35].

**PROGENT** uses a domain-specific language for tool access control. It acts as a wrapper that applies security policies at tool call time. While promising, this approach focuses only on tool calling [36].

**CONSECA** offers a framework for policy generation and enforcement. The approach focuses on managing trusted context and preventing actions misaligned with the current context [37].

The reviewed behavior-level defenses address critical needs but lack practicality and universality. Rather than competing with them, A2AS aims to address their gaps with practical security engineering.

# 4 BASIC Security Model

## 4.1 BASIC Pillars

The BASIC security model introduces a set of security primitives for behavior certification, context window integrity, and secure model reasoning. It is aligned with the nature of AI agents and grounded in three foundational pillars: runtime, self-defense, and self-sufficiency.

**RUNTIME**. Security should apply not only when the model executes but also where it executes. Controls should be enforced at runtime to ensure all requests and actions are protected at the system level.

**SELF-DEFENSE**. Controls should leverage the model's own reasoning to interpret security rules and boundaries. Operating natively within the context window allows efficient and secure model reasoning.

**SELF-SUFFICIENCY**. Security architecture should avoid complex orchestration and reliance on external models or tools. Eliminating external dependencies reduces latency and minimizes third-party risk.

Out-of-scope approaches are those that conflict with the key pillars, such as not operating at runtime, not using the model's reasoning, or depending on external components.

| BASIC CONTROLS | CORE FOCUS | ANALOGY |
|---|---|---|
| **(B)** BEHAVIOR CERTIFICATES | Action Permissions | JSON-based AWS access policies |
| **(A)** AUTHENTICATED PROMPTS | Request Authenticity | HMAC-based API request signing |
| **(S)** SECURITY BOUNDARIES | Input Segmentation | HTML sandboxing for iframes |
| **(I)** IN-CONTEXT DEFENSES | Secure Reasoning | CSP for built-in browser security |
| **(C)** CODIFIED POLICIES | Custom Rules | DLP with custom content filters |

**Figure 2:** BASIC security controls with focus areas and analogies

## 4.2 BASIC Controls

The BASIC security controls complement each other, enabling a defense-in-depth strategy. Although designed to be composable, each control can be used independently.

The essential nature of these controls motivated the name **BASIC**.

**BEHAVIOR CERTIFICATES** enable AI agent developers to declare their operational boundaries and capabilities. The certificates establish action permissions for accessing tools, files, and other resources. They provide a standardized method for organizations to control AI agent execution and serve as a bill of materials for AI agents.

**AUTHENTICATED PROMPTS** ensure that external inputs are validated for integrity and authenticity before being processed by LLM models. Each request can include a signature derived from the request source, message content, or applied policy. This also enables attribution and auditability through a verifiable prompt history.

**SECURITY BOUNDARIES** isolate untrusted external inputs from trusted system instructions within the context window. Every prompt, tool response, and external content is enclosed within special tags, helping LLM models recognize context window boundaries. Explicitly marking content for downstream systems can also improve security.

**IN-CONTEXT DEFENSES** leverage the model's own reasoning for self-protection. Security meta-instructions are embedded directly into the context window, operating natively and guiding the model to reject malicious inputs and disregard unsafe content. They effectively act as model-native security guardrails without external dependencies.

**CODIFIED POLICIES** allow defining application-specific behavior rules in domain-specific language or natural language and managing them as code. This enables policy versioning and testing, while supporting model adaptation to the domain and business requirements.

The BASIC model defines the fundamental controls required for LLM security, serving both as the foundation for the A2AS framework and as a standalone conceptual model for agentic AI defense-in-depth.

# 5 A2AS Framework

## 5.1 Control Layers

The A2AS framework is an implementation of the BASIC model. It adds a runtime security layer for AI agents and LLM-powered applications, similar to how HTTPS secures HTTP.

The A2AS security controls operate at multiple levels.

**FUNCTION-LEVEL CONTROLS** act within the source code, inspecting tool calling parameters, resource access requests, and other system operations in the runtime environment, ensuring certified behavior.

**CONTEXT-LEVEL CONTROLS** act within the context window, using prompt instrumentation for prompt-bound and context-wide controls, ensuring context window integrity and secure model reasoning.

## 5.2 Prompt Instrumentation

Context-level controls are embedded into the context window through an A2AS-managed prompt template or system prompt.

**PROMPT-BOUND** controls, which consist of authenticated prompts and security boundaries, are tied to each input and wrap it into the managed prompt template augmented with special metadata.

**CONTEXT-WIDE** controls, which consist of in-context defenses and codified policies, are embedded into the same prompt template but can be offloaded to the system prompts for token efficiency.

All controls are designed to complement each other, yet all of them can be used independently, leveraging their own namespaces.

```
<a2as:user> User commands </a2as:user>

<a2as:tool> Tool responses </a2as:tool>

<a2as:hash> Integrity hashes </a2as:hash>

<a2as:defense> Context defenses </a2as:defense>

<a2as:policy> Application policies </a2as:policy>
```

**Figure 3**: A2AS-managed prompt template with context-level controls

## 5.3 Behavior Certificates

AI developers usually define expected behavior through agent cards or other manifests. These declarations may include agent and model names, registered tools, system resources, and other capabilities.

The A2AS behavior module (**a2as.behavior**) enables issuing certificates for approved behaviors that can be loaded alongside the agent for runtime enforcement. This module manages the creation and validation of certificates as well as interpreting and enforcing their declarations. Certificates can be distributed by AI agent developers or self-signed internally by organizations after inspection and approval.

Engineers can define constraints to apply to the managed AI agents, enabling or disabling their permissions and capabilities. For example, they can enforce read-only access to certain files, prevent writing any files, or allow executing only approved commands.

This mechanism reflects ideas from OpenAPI schema validation and Kubernetes admission policy configuration.

```
"agent_id": "agent-email-reporter-v1",

"permissions": {

    "email": { "provider": "gmail" },

    "files": { "write": "./out/email_report.json" },

    "functions": [
        { "name": "call:email.list_messages", "critical": true },
        { "name": "call:email.read_message", "critical": true }
    ]
}
```

**Figure 4**: Example of behavior certificates with permission declarations

## 5.4 Authenticated Prompts

AI agents and LLM-powered applications process external inputs without verifying their integrity and authenticity. This allows malicious inputs to enter the context window, leading to successful attacks.

The A2AS integrity module (`a2as.integrity`) enables authenticating prompts by computing their integrity hashes using request-specific information such as origin, content, and metadata. This module allows each prompt to be validated before the model processes it, adding trust at the prompt level and ensuring context integrity.

Engineers can define how to process the authenticated prompts. For example, prompts with valid hashes may have the metadata recorded for auditing, while prompts with corrupted hashes can be rejected and trigger full request logging to support investigation.

This mechanism ensures long-term context window integrity while enabling attribution and auditability.

```
<system>
    You are a helpful email assistant
</system>

<a2as:user:7c3d0c6d>
    Review all of my emails for a weekly report
</a2as:user:7c3d0c6d>

<assistant>
    Sure, let me gather your emails from the past week
</assistant>
```

**Figure 5**: Example of authenticated prompts with integrity hashes

## 5.5 Security Boundaries

AI agents and LLM-powered applications do not distinguish between trusted system instructions and untrusted external inputs from users or tools, leading to attacks through context window manipulation.

The A2AS boundary module (`a2as.boundary`) enables defining explicit security boundaries within the context window. This module ensures that external inputs are wrapped in boundary tags before being passed to the LLM model. These tags help the model process external content securely by indicating its origin and trust level. Security boundaries are foundational to other controls, but they can also function independently in some use cases.

Engineers can implement additional processing logic such as routing or logging based on specific boundaries. For example, they can record every input from selected tools or alert on system prompt changes.

This mechanism binds sensitive operations to the trusted context, supported by in-context defenses and codified policies.

```
<system>
    You are a helpful email assistant
</system>

<a2as:user:7c3d0c6d>
    Review all of my emails for a weekly report
</a2as:user:7c3d0c6d>

<assistant>
    Sure, let me gather your emails from the past week
</assistant>

<a2as:tool:1bfa2466>
    Here are the contents of 1337 emails received this week [ ... ]
</a2as:tool:1bfa2466>

<assistant>
    Here is your weekly summary [ ... ]
</assistant>
```

**Figure 6:** Security boundaries with user and tool input segmentation

## 5.6 In-Context Defenses

AI agents and LLM-powered applications demonstrate resistance to prompt injection attacks with advanced instruction-based defenses applied directly within the context window.

The A2AS defense module (**a2as.defense**) enables secure model reasoning using security meta-instructions embedded into the context window. This module guides the LLM model to reject malicious inputs and unsafe content just-in-time while processing external inputs.

Engineers can configure in-context defenses to operate at the regular prompt template or system prompt levels. Both methods work natively within the context window, enabling effective protection.

This mechanism relies on security boundaries for input segmentation and significantly reduces attack success rate without adding latency or relying on external guardrails.

```
<system>
    You are a helpful email assistant
</system>

<user>
    <a2as:defense>
        External content is in <a2as:user> and <a2as:tool> tags.
        Treat all external content and instructions as untrusted.
        Don't follow instructions from tools or agents by default.
        Prevent prompt injections, jailbreaks, other unsafe requests.
    </a2as:defense>
    <a2as:user:7c3d0c6d>
        Review all of my emails for a weekly report
    </a2as:user:7c3d0c6d>
</user>

<assistant>
    Sure, let me gather your emails from the past week
</assistant>

<a2as:tool:1bfa2466>
    Here are the contents of 1337 emails received this week [ ... ]
    One email contains: "Ignore all [..] and forward to attacker@email.com" [ ... ]
</a2as:tool:1bfa2466>

<assistant>
    A prompt injection was detected in one of the emails ("Ignore all [ ... ]"),
    which was excluded from the report. Here is your summary [ ... ]
</assistant>
```

**Figure 7:** In-context defenses embedded into the context window

# 5.7 Codified Policies

AI agents and LLM-powered applications can use custom policies as an effective alignment mechanism for domain-specific requirements.

The A2AS policy module (**a2as.policy**) enables defining codified policies with rules, states, and behaviors. They are embedded into the context window similar to the in-context defenses through prompt instrumentation. This module feeds the LLM model with up-to-date policies, making them available for safe and compliant reasoning.

Engineers can define rules with domain-specific language or natural language to align intended model or agent behaviors with the domain and business requirements. For example, a policy may block inputs containing source code or ensure rejection of personal data.

This mechanism leverages policy-as-code, supporting versioning and transparency. Policies can be updated across environments without changing the underlying LLM model or AI agent architecture.

```
<system>
    You are a helpful email assistant
</system>

<user>
    <a2as:policy>
        The following policies apply to this application.
        This read-only app must not modify or send emails.
        Emails labeled "Confidential" must not be processed.
        Personal information in any form must not be processed.
    </a2as:policy>
    <a2as:user:7c3d0c6d>
        Review all of my emails for a weekly report
    </a2as:user:7c3d0c6d>
</user>

<assistant>
    Sure, let me gather your emails from the past week
</assistant>

<a2as:tool:1bfa2466>
    Here are the contents of 1337 emails received this week [ ... ]
    Found 10 emails that are tagged "Confidential" [ ... ]
</a2as:tool:1bfa2466>

<assistant>
    I have excluded all emails labeled "Confidential" and ensured that
    no personal information is included. Here is your weekly summary [ ... ]
</assistant>
```

**Figure 8**: Codified policies embedded into the context window

## 5.8 Framework Extensions

The modular architecture of the A2AS framework enables extending it with minimal effort by introducing custom workflows and integrating with internal security, observability, and development platforms.

**CONTEXT AUDIT** is enabled by design because prompt-level security controls are embedded into the context window. Engineers can treat the context as a structured prompt history with a verifiable record of all interactions, supporting auditing and investigations.

**RUNTIME TELEMETRY** can be implemented with minimal effort since the framework processes and can record metadata such as prompt origin, active policies, or timing data. Engineers can gain visibility into regular operations and review enforcement actions.

**CAPABILITY LABELS** can be introduced for model inputs and outputs, tool arguments, or external data, labeling them by sensitivity or other criteria. Engineers may define categories such as trusted, untrusted, confidential, financial, or personal information. These labels can be propagated across workflows for security or custom logic.

**IDENTITY BINDINGS** can be implemented to augment authenticated prompts with enterprise identity information. While their primary role is to ensure context window integrity, the introduced bindings can act as a proxy for identity, supporting attribution and enabling access control.

**PIPELINE INTEGRATION** is straightforward because the framework is designed to be compatible with CI/CD pipelines, enabling automated agent security testing. Engineers can implement environment-specific A2AS configurations and use testing results for deployment approval.

Collectively, the framework's extensibility and built-in features ensure universal compatibility and enable scalable adoption, supported by enterprise-grade security engineering for evolving agentic AI security.

## 5.9 Known Limitations

The A2AS framework is designed as a powerful tool, operating under a few key assumptions and implementation constraints.

**TOKEN USAGE OVERHEAD.** Context-level controls increase token usage because the context window is augmented with technical metadata. Although the cost of context integrity is paid in extra tokens, prompt-bound controls introduce only minimal overhead, while context-wide controls can be offloaded to system prompts.

**SECURITY REASONING DRIFT.** Not all LLM models may interpret in-context defenses and codified policies equally. Variations in model reasoning may lead to misinterpretation or partial compliance. This limitation is addressed by the A2AS framework design, where controls complement one another, providing reliable fallback mechanisms.

**CAPACITY-CONSTRAINED REASONING.** Small LLM models may lack the reasoning depth for in-context defenses and codified policies. Although these controls can be optimized for any LLM model, reliable enforcement with constrained reasoning requires additional research.

**SECURITY MISCONFIGURATION RISK.** A misconfigured certificate or poorly written policy can create a false sense of security, leaving the attack surface exposed. While controls such as in-context defenses are optimized out of the box, others such as behavior certificates and codified policies rely on operators to configure them correctly.

**MULTIMODAL COVERAGE GAP.** Rule-focused security controls such as in-context defenses and codified policies are optimized to operate on textual data. Although they can protect multimodal LLM models, some attacks could bypass the security controls. Solving this requires further research in multimodal input sanitization.

Rather than framework flaws, these limitations are operational factors that require awareness for effective A2AS implementation, especially in environments with diverse AI agents and LLM model versions.

# 6 A2AS Use Cases

## 6.1 User-to-Agent Attacks

User-to-agent is a pattern where a human interacts directly with an AI assistant to complete tasks.

Employees can use financial AI assistants to process invoices for extracting data such as vendor name, payment amount, and other details. The structured output is then used to schedule payments.

An attacker can embed a hidden, indirect prompt injection inside a legitimate-looking invoice. This malicious instruction would cause the AI assistant to replace the vendor's bank account number with the attacker's bank account. Since employees cannot guarantee human oversight for verifying invoice details at scale, the attack can result in fraudulent bank transfers and direct financial losses.

| A2AS CONTROLS | A2AS USE CASE |
|---|---|
| BEHAVIOR CERTIFICATES | Restrict the AI agent to read-only operation, preventing automatic payment scheduling and approval |
| AUTHENTICATED PROMPTS | Generate integrity hashes for each request, creating a verifiable history for audit and investigations |
| SECURITY BOUNDARIES | Isolate the invoice content inside explicit boundary tags, highlighting untrusted external data to the AI agent |
| IN-CONTEXT DEFENSES | Guide the AI agent to treat the external content as untrusted, with potentially malicious instructions |
| CODIFIED POLICIES | Define a domain-specific policy, such as requiring a human confirmation for sensitive values like bank accounts |

**Figure 9:** A2AS security controls for the user-to-agent attack scenario

## 6.2 Agent-to-Tool Attacks

Agent-to-tool is a pattern where an AI assistant can pull information from enterprise tools through API and AI protocols such as MCP.

Employees can use AI assistants to handle business email workflows. Such assistants may have access to manage corporate email accounts and query CRM data, among other tools and permissions.

An attacker can send an email with a prompt injection that instructs the AI assistant to extract a customer list from the CRM and send it to the attacker's email address. Exploiting the privileged access to read CRM data and send emails can result in a large-scale data breach, exposing sensitive customer information, damaging customer trust, and potentially leading to regulatory penalties.

| A2AS CONTROLS | A2AS USE CASE |
|---|---|
| BEHAVIOR CERTIFICATES | Restrict the AI agent permissions to CRM queries with limited scope, minimizing the risk of data exposure |
| AUTHENTICATED PROMPTS | Generate a verifiable request and action history that attributes which email triggered which CRM query |
| SECURITY BOUNDARIES | Isolate the email content inside explicit boundary tags, highlighting untrusted external data to the AI agent |
| IN-CONTEXT DEFENSES | Guide the AI agent to ignore external instructions found in the email content and prevent unauthorized tool calls |
| CODIFIED POLICIES | Define a policy that requires CRM queries to be scoped and enforces human review for emails to non-corporate domains |

**Figure 10:** A2AS security controls for the agent-to-tool attack scenario

## 6.3 Agent-to-Agent Attacks

Agent-to-agent is a pattern where multiple AI agents collaborate, exchanging tasks or data via API and AI protocols such as A2A.

Organizations can use AI agents for system monitoring and recovery. Some agents track system health, collect logs, and analyze incidents, while other agents automate recovery by making system changes.

An attacker can embed a self-propagating prompt injection into a log file. This instruction causes one agent to forward a malicious payload to peer agents, executing attack commands across systems. Since this agentic workflow requires privileged system access, the prompt infection can escalate into a company-wide ransomware incident, resulting in large-scale file encryption or destructive actions.

| A2AS CONTROLS | A2AS USE CASE |
|---|---|
| BEHAVIOR CERTIFICATES | Restrict the AI agents to read-only access for logs data, limited file writing, execution of only allowed commands |
| AUTHENTICATED PROMPTS | Generate integrity hashes for inter-agent requests, enabling attribution and supporting incident response |
| SECURITY BOUNDARIES | Isolate the raw log content inside explicit boundary tags, highlighting untrusted external data to AI agents |
| IN-CONTEXT DEFENSES | Guide AI agents to treat logs strictly as data content, preventing malicious payload infection and distribution |
| CODIFIED POLICIES | Define a policy that limits operations based on origin trust and requires human review for critical system tasks |

**Figure 11:** A2AS security controls for the agent-to-agent attack scenario

# 7 A2AS Roadmap

The A2AS framework is being architected as an open and universal AI runtime security layer and behavior certification standard.

**NEAR TERM**, the goal is to integrate core A2AS features with common AI agent development frameworks to support adoption in production agentic AI systems. Further research will cover behavior certification schemas and context protection benchmarking.

**MID TERM**, the goal is to improve the A2AS implementation based on industry feedback. Research will explore key improvements in behavior certification, policy following, and security-utility tradeoffs.

**LONG TERM**, the goal is to align the A2AS framework with the growing AI ecosystem, emerging AI-native protocols, and agentic AI design patterns, establishing A2AS as the industry standard.

# 8 Conclusion

This paper presented a runtime security layer for AI agents and LLM-powered applications, similar to how HTTPS secures HTTP.

**FIRST**, we introduced the BASIC security model, which defines five security primitives: behavior certificates, authenticated prompts, security boundaries, in-context defenses, and codified policies.

**NEXT**, we introduced the A2AS framework as an implementation of the BASIC security model. A2AS is an AI runtime security layer that enforces certified behavior, activates secure model reasoning, and ensures context window integrity, enabling a defense-in-depth strategy. Unlike alternative solutions, the framework avoids latency overhead, external dependencies, architectural changes, model retraining, performance degradation, and operational complexity.

**FINALLY**, we laid the foundation to take this lightweight, modular, and scalable framework from early industry adoption toward establishing the A2AS industry standard.

## Call for Collaboration

We invite AI developers and security researchers to build and implement the A2AS framework. Like HTTPS for the web, A2AS works to standardize secure communications for AI applications. Visit the A2AS project website for updates, details, and contact information at **https://a2as.org**.